\documentclass[12pt]{article}
\usepackage{amssymb}

\def \square {\hbox{$\sqcup\!\!\!\!\sqcap$}} 
\newcommand{\be}{\begin{equation}}
\newcommand{\ee}{\end{equation}} 
\newcommand{\bea}{\begin{eqnarray}}
\newcommand{\eea}{\end{eqnarray}}

\catcode`\@=11 
\@addtoreset{equation}{section}
 
\catcode`\@=11

\begin{document}

\begin{titlepage}

\begin{flushright} 
{\tt FTUV/99-67\\ 
     IFIC/99-70\\ 
     hep-th/9910076}
 \end{flushright}

\bigskip

\begin{center}

{\bf \LARGE AdS$_2$/CFT$_1$ correspondence and near-extremal 
black hole entropy}

\bigskip 

 J.~Navarro-Salas\footnote{\sc jnavarro@lie.uv.es} and
 P.~Navarro\footnote{\sc pnavarro@lie.uv.es}.

\end{center}

\bigskip

\begin{center}
\footnotesize
        Departamento de F\'{\i}sica Te\'orica and 
	IFIC, Centro Mixto Universidad de Valencia-CSIC.
	Facultad de F\'{\i}sica, Universidad de Valencia,	
        Burjassot-46100, Valencia, Spain. 
\end{center}               

\normalsize

\bigskip
\bigskip

\begin{center}
			{\bf Abstract}
\end{center}
We provide a realization of the AdS$_2$/CFT$_1$ correspondence in terms of asymptotic
symmetries of the AdS$_2\times$S$^1$ and AdS$_2\times$S$^2$ geometries arising in 
near-extremal BTZ and Reissner-Nordstr\"om   
black holes. Cardy's formula exactly accounts for the deviation
of the Bekenstein-Hawking entropy from extremality. We also argue that this result can
be extended to more general black holes near extremality.
\end{titlepage}

\newpage

\section{Introduction}

  Since the discovery of the thermodynamical properties of black holes a crucial open
problem has been to find a microscopical structure responsible for the
Bekenstein-Hawking entropy. In the last few years this question has started to receive
some answers. The discovery of D-branes led to an explicit statistical derivation of
the black hole entropy for extremal \cite{StrVafa} and near-extremal \cite{CalMald}
black holes (see also the reviews \cite{M1,P}). In a different context, Strominger has
proposed \cite{Strominger} a unified way to account for the Bekenstein-Hawking
entropy of black holes whose near-horizon geometries are locally similar to the
BTZ black holes \cite{BTZ}. The idea of the approach of \cite{Strominger} is that a
conformal symmetry of the gravity theory can control the asymptotic density of
states, irrespective of the details of quantum gravity theory, thus providing a
statistical explanation to the area formula for the entropy and, in turn, a sort of
universality. Strominger's argument is based in the holographic relation, first
discovered by Brown and Henneaux \cite{BH}, between gravity on AdS$_3$ and a
two-dimensional conformal field theory on the boundary. AdS$_3$ gravity possesses a set
of asymptotic symmetries closing down two copies of the Virasoro algebra with central
charge $c={3\ell \over 2G}$, where G is Newton's constant and $-1/\ell^2$ is the
cosmological constant. Using Cardy's formula \cite{Cardy}
 for the boundary CFT$_2$ one reproduces
the expected entropy. However, the validity of Cardy's formula requires that the lowest
eigenvalues of the Virasoro operators $L_0$ and $\bar{L}_0$ vanish. As it has been
pointed out in \cite{Carlip2}, this is not the case of the boundary theory of AdS$_3$
gravity, because it is, up to global issues, Liouville theory \cite{CHvD}. The
asymptotic level density of states is then controlled by the effective central charge
\cite{KS} which, for Liouville theory, turns out to be equal to one and therefore
cannot properly account for the entropy. However, the fact that the 
entropy fits Cardy's formula with the ordinary central charge seems to indicate that
gravity theory itself can provide relevant information about the microscopic theory,
but apparently not enough to characterise it completely. Interesting attempts to avoid
the restrictions of 2+1 dimensions to explain the Bekenstein-Hawking entropy by means
of symmetry principles has been given in \cite{Carlip3, Carlip4, Solod} by considering
the horizon as a boundary.

  Among the family of AdS$_D$/CFT$_{D-1}$ dualities \cite{Maldacena},
the pure gravity case AdS$_3$/CFT$_2$ is the best understood. 
In contrast, the AdS/CFT correspondence
in two space-time dimensions is quite enigmatic. Some progress has been made
in \cite{Strominger2,MMStr,CadMig,GibbTown}.
One of the aims of this paper is to further investigate the AdS$_2$/CFT$_1$ 
correspondence in terms of asymptotic symmetries.
In section 2 we shall analyse the relation between the first sub-leading terms
in the asymptotic expansion of the metric field,
obeying suitable AdS$_2$ boundary conditions, and the stress tensor of the
boundary theory, as happen in higher dimensional situations \cite{Nav,Balasu}.
Following a similar line of reasoning as in \cite{Strominger} we shall show that
the application of Cardy's formula to the unique
copy of the Virasoro algebra emerging as an asymptotic symmetry, yields
to the entropy of spinless BTZ black holes. But more interestingly,
the AdS$_2$/CFT$_1$ correspondence, implemented via asymptotic symmetries, 
can be used to
correctly account for the deviation of the
Bekenstein-Hawking entropy from extremality in the near-horizon approximation.
On general grounds, two-dimensional Anti-de Sitter space naturally arises in
the near-horizon limit around the degenerate radius of coincident horizons
\cite{CruzNavFab}. Therefore, a way to study Maldacena's duality in D=2 and its 
implication for black holes is to consider gravity 
theories having black hole solutions with degenerate horizons.
In section 3 we consider near-extremal BTZ
black holes and in section 4 four-dimensional
Reissner-Nordstr\"om black holes near extremality. Finally, in section 5
we show that the above results can be extended to any black hole
with degenerate horizons that can be properly described by a two-dimensional
effective theory.

\section{Dimensional reductions of AdS$_3$ gravity and the AdS$_2$/CFT$_1$
correspondence}

  Einstein gravity on AdS$_3$ is described by the action
\be
S={1 \over 16\pi G}\int d^3x\sqrt{-g}(R+{2 \over \ell^2})
\ee
and we can dimensionally reduce the theory \cite{AchuO,LMK}
via a decomposition of the
metric of the form
\be
ds^2_{(3)} = g_{\mu\nu}dx^{\mu}dx^{\nu}+\ell^2\phi^2(x)
(d\theta + A_{\mu}(x)dx^{\mu})^2\, , \qquad \mu ,\nu =0,1 
\ee
where $ds^2=g_{\mu\nu}dx^{\mu}dx^{\nu}$ is a two-dimensional metric, $\phi$ a
scalar (dilaton) field and $A_{\mu}$ a Kaluza-Klein ${\cal U}(1)$ gauge field.
The two-dimensional effective theory is governed by the action
\be
{\ell \over 8G}\int d^2x\sqrt{-g}\phi(R+{2 \over \ell^2}
-{\ell^2 \over 4}\phi^2F_{\mu\nu}F^{\mu\nu}) \, . \label{action2}
\ee
The equations of motion of the gauge field imply that
\be
2\frac{\ell^3 \phi^3}{\sqrt{-g}}F_{+-}=constant \, , \label{constant}
\ee
where $x^{\pm}=x^0\pm x^1$ and $F_{+-}=\partial_+A_- -\partial_-A_+$.
  Moreover, by varying the dilaton one obtains
\be
R+{2 \over \ell^2}-{3 \over 4}\ell^2\phi^2F^2=0 \, ,
\ee  
and using (\ref{constant}) one gets
\be
R=-{2 \over \ell^2}-{3 \over 2}\frac{J^2}{\ell^2\phi^4} \, , \label{eqR}
\ee
where J is related with the integration constant of (\ref{constant}).
The action (\ref{action2}) turns out to be then
\be
{\ell \over 8G}\int d^2x\sqrt{-g}(\phi R+V(\phi)) \, , \label{action3}
\ee
where
\be
V(\phi)=\phi\left({2 \over \ell^2}-\frac{J^2}{2\ell^2\phi^4}\right) \, .
\label{potential}
\ee
The most general solution of (\ref{action3}) with a linear dilaton corresponds
to the dimensional reduction of the BTZ black hole which, in the Schwarzschild
gauge, takes the form (with $A_t = -{4GJ \over r^2}$)
\bea 
ds^2 &=& -\left({r^2 \over \ell^2}-8GM+{16G^2J^2 \over r^2}\right)dt^2+
\frac{dr^2}{\left({r^2 \over \ell^2}-8GM+{16G^2J^2 \over r^2}\right)} \, , 
\label{metric0} \\
\phi &=& {r \over \ell} \, . \label{dilaton0}
\eea
  The two event horizons are located at
\be
r_{\pm}^2=4GM\ell^2\left( 1\pm\sqrt{1-\left({J \over M\ell}\right)^2}\right)
\ee
and the outer horizon give rise to the entropy \cite{LMK}
\be
S={2\pi r_+ \over 4G}=\pi\sqrt{{\ell(\ell M +J) \over 2G}}+
\pi\sqrt{{\ell(\ell M-J) \over 2G}} \, , \label{entropy}
\ee
which reproduce as expected the entropy of the original three-dimensional
theory. To get a two-dimensional AdS geometry from (\ref{eqR}) we have
two different ways. We can restrict the theory to the spinless sector $J=0$
with
\be
R=-{2 \over \ell^2} \, ,
\ee
or we can fix the value of the dilaton in a way consistent with the equation
of motion \cite{CruzNavFab}
\be
\square\phi = V(\phi) \, ,
\ee
where $V(\phi)$ is given by (\ref{potential}). This implies that the value
of the dilaton $\phi=\phi_0$ should be a zero of the potential
\be
V(\phi_0)=\phi_0\left({2 \over \ell^2}-{J^2 \over 2\ell^2\phi_0^4}\right) =0 \, ,
\ee
and then
\be
R=-V'(\phi_0)=-{8 \over \ell^2} \, .
\ee

  In the remaining part of this section we shall study the first possibility,
which is equivalent to consider the Jackiw-Teitelboim model of two-dimensional
gravity \cite{Jackiw}. The second way to get an AdS$_2$ geometry
(AdS$_2\times$S$^1$ from the three-dimensional point of view) will be
widely analysed in the next section, although the basic features of the
AdS$_2$/CFT$_1$ correspondence considered in our approach will be presented
here.

 The reduced theory with $J=0$ coincides with the Jackiw-Teitelboim model
\be
S={\ell \over 8G}\int d^2x\sqrt{-g}\phi(R+{2 \over \ell^2}) \, ,
\ee
whose solutions are of the form
\bea
ds^2 &=& -({x^2 \over \ell^2}-a^2)dt^2+({x^2 \over \ell^2}-a^2)^{-1}dx^2 \, ,
\label{metric1} \\
\phi &=& {x \over \ell} \, ,
\eea
with $a^2=8GM$.
The metrics (\ref{metric1}) are locally AdS$_2$ and in order to define a  
quantum theory we have to specify boundary conditions for the fields at
infinity. Mimicking the analysis of three-dimensional gravity \cite{Nav}
we shall assume the following asymptotic behaviour of the metric 
\footnote{These boundary conditions where first introduced in \cite{CadMig}.}
\bea
g_{tt} &=& -{x^2 \over \ell^2}+\gamma_{tt}(t)+{\cal O}({1 \over x^2}) \, , \\
g_{tx} &=& {\gamma_{tx} \over x^3}+{\cal O}({1 \over x^5}) \, , \\
g_{xx} &=& {\ell^2 \over x^2}+{\gamma_{xx} \over x^4}+{\cal O}({1 \over x^6}) \, ,
\eea
where we have now introduced the first sub-leading terms in the expansion
aiming to relate them with a conformal field on the boundary.

  The infinitesimal diffeomorphisms $\zeta^a(x,t)$ preserving the above
boundary conditions are
\bea
\zeta^t &=& \epsilon (t)-{\ell^4 \over 2x^2}\epsilon''(t)
+{\cal O}({1 \over x^4}) \label{dif1} \, , \\
\zeta^x &=& -x\epsilon'(t)+{\cal O}({1 \over x}) \, . \label{dif2}
\eea
Using the "gauge" diffeomorphisms
\bea
\zeta^t={\alpha^t(t) \over x^4}+{\cal O}({1 \over x^5}) \, ,\\
\zeta^x={\alpha^x(t) \over x}+{\cal O}({1 \over x^2}) \, ,
\eea
where $\alpha^t$ and $\alpha^x$ are arbitrary functions, one can easily show
that the only gauge invariant quantity is
\be
\Theta_{tt}=\kappa(\gamma_{tt}-{\gamma_{xx} \over 2\ell^4}) \, ,
\ee
where $\kappa$ is a constant coefficient.

The action of the infinitesimal diffeomorphism (\ref{dif1}-\ref{dif2}) on the
metric induces the following transformation for the function $\Theta_{tt}$:
\be
\delta_{\epsilon}\Theta_{tt}=\epsilon (t)\Theta_{tt}'
+2\Theta_{tt}\epsilon'(t)-\kappa\ell^2\epsilon'''(t) \, . \label{law}
\ee
So, the quantity $\Theta_{tt}$ behaves as the (chiral component of the)
stress tensor of a conformal field theory.
To evaluate the central charge we need to know the coefficient $\kappa$. To this
end we have to work out the Noether charges associated to the above asymptotic
symmetries. Using the decomposition of the metric
\be
ds^2=-N^2dt^2+\sigma^2(dx+N^xdt)^2 \, ,
\ee
the bulk Hamiltonian of the theory is given by
\be
H_0=\int dx(N{\cal H}+N^x{\cal H}_x) \, ,
\ee
where the constraints are
\bea
{\cal H} &=& - \Pi_{\phi}\Pi_{\sigma}+({\phi' \over \sigma})'
-{\sigma\phi \over \ell^2} \, , \\
{\cal H}_x &=& \Pi_{\phi}\phi'-\sigma\pi_{\sigma}' \, ,
\eea
and the momenta
\bea
\Pi_{\phi} &=& N^{-1}(\sigma^{-1}+(N^x\sigma)') \, , \\
\Pi_{\sigma} &=& N^{-1}(-\dot{\phi}+N^x\phi') \, .
\eea
  The full Hamiltonian is given by
\be
H=H_0+K \, ,
\ee
where $K$ is a boundary term necessary to have well-defined variational
derivatives. Assuming the boundary condition for the dilaton
\be
\phi = {x \over \ell} + {\ell \over 2x}\gamma_{\phi\phi}(t)
+{\cal O}({1 \over x^2})
\ee
and imposing that $K$ vanishes for $a^2=0$, the boundary term $K$ can
be worked out \cite{CadMig}
\be
K(\epsilon)={\ell \over 4G}\lim_{x\to\infty}
\left\{ -{x \over \ell}\zeta^{\bot}(\phi'-{1 \over \ell}) 
+ {x \over \ell}\frac{\partial\zeta^{\bot}}{\partial x}(\phi-{x \over \ell})
+ {x^3 \over 2\ell^4}\zeta^{\bot}(g_{xx}-{\ell^2 \over x^2})
+ {\ell \over x}\zeta^{\Vert}\Pi_{\sigma}\right\} \, .
\ee
  Using the asymptotic expansion for the metric and the dilaton we obtain
\be
K(\epsilon) = {\epsilon \over 4G}\left({1 \over 2\ell^4} 
\gamma_{xx}+\gamma_{\phi\phi}\right) \, .
\ee
Moreover, the equation for the dilaton
\be
\square\phi = {2 \over \ell^2}\phi
\ee
allows to relate $\gamma_{\phi\phi}$ with the remaining quantities
\be
\gamma_{\phi\phi} = \left(\gamma_{tt}
-{\gamma_{xx} \over \ell^4}\right) \, ,
\ee
and then $K(\epsilon)$ can be written in terms of the unique gauge 
invariant quantity
\be
K(\epsilon) = \epsilon{1 \over 4G}\left(\gamma_{tt}-{1 \over 2\ell^4}
\gamma_{xx}\right) \, .
\ee 

  The standard identification of $K(\epsilon)$ in terms of the stress tensor
\cite{CFT}
\be
K(\epsilon) = \epsilon\Theta_{tt}
\ee
allows us to know the coefficient $\kappa$, which turns out to be
\be
\kappa = {1 \over 4G} \, .
\ee
  We still have to compute the central charge. Defining the Fourier components
$L_n^R$ of $\Theta_{tt}$ as
\be
L_n^R = {1 \over 2\pi\ell}\int_0^{2\pi\ell}dt\Theta_{tt}\ell e^{int/\ell} \, ,
\ee
where we assume periodicity of $t$ in the interval $0\leq t<2\pi\ell$, the
Poisson algebra can be expressed as follows
\be
\{ L_n^R, L_m^R\} = \delta_{\epsilon_m}L_n^R \, ,
\ee
where $\epsilon_m = \ell e^{imt/\ell}$. Using (\ref{law}) it is easy to get the
following Virasoro algebra
\be
i\{ L_n^R, L_m^R \} = (n-m)L_{n+m}^R + {c \over 12}n^3\delta_{n,-m}
\ee
with central charge
\be
c = 12\kappa\ell = {3\ell \over G} \label{centralc} \, .
\ee
For the black holes
(\ref{metric1}) we have a constant value of $\Theta_{tt}$
\be
\Theta_{tt}= {\kappa \over 2}a^2 = {1 \over 8G}a^2
\ee
and, in terms of the mass $a^2=8GM$, we have
\be
L_0^R= \ell\Theta_{tt} = M \ell \label{L0}
\ee
Observe that to have a Virasoro algebra of the Neveu-Schwartz form we must
shift the Ramond-type generator: $L_0^R\rightarrow 
L_0^{NS}=L_0^R+{c \over 24}$.

  The results (\ref{centralc}) and (\ref{L0}) allows us to compute the
asymptotic density of states using Cardy's formula
\be
\log\rho(\Delta)\sim 2\pi\sqrt{c\Delta \over 6} \, ,
\ee
where $\Delta$ is the eigenvalue of the Virasoro generator $L_0^{NS}$. In our
case $c={3\ell \over G}$ and $\Delta =M\ell+{\ell \over 8G}$. For large mass
$\Delta\gg c$ we get the following statistical entropy
\be
S = 2\pi\sqrt{M\ell^2 \over 2G} \, ,
\ee
which coincides with the thermodynamical formula (\ref{entropy}) with $J=0$.
We should stress the fact that, in contrast with the analysis of 
\cite{CadMig}, which uses the convention $4G=\ell$,
we have found an exact agreement between the statistical
entropy of the two-dimensional black hole (\ref{metric1}) and the
corresponding 2D Bekenstein-Hawking formula. The discrepancy comes from the
evaluation of the central charge. Our result is $c={3\ell \over G}$ and the
authors of \cite{CadMig} claim that $c=24({\ell \over 4G})$.

We must stress now the important fact that the statistical entropy is independent of
the length of the interval of the compactified parameter $t$. If we choose a different
periodicity for $t$: $0<t<2\pi\beta$, the central charge shift 
$c\rightarrow c{\ell \over \beta}$, but $L_0^R$ get modified 
$L_0^R\rightarrow {\beta \over \ell}L_0^R$ in such a way that $cL_0^R$, and hence
the entropy, is not sensitive to the compactification scale.

\section{Near-extremal BTZ black holes}

  The second possibility to get a AdS$_2$ geometry 
in AdS$_3$ gravity is by means of a constant
dilaton solution. This can be obtained performing a perturbation around the 
degenerate radius of the extremal solutions,
keeping the angular momentum $\mid{J \over \ell}\mid = M_0$ fixed,
(see \cite{CruzNavFab} for the general case):
\be
M = \mid J/\ell\mid(1+k\alpha^2) \, ,
\ee
where $\alpha$ is an infinitesimal parameter $0<\alpha\ll 1$ and $k$ is an arbitrary
positive constant.
Introducing the coordinates $(\tilde{t},\tilde{x})$
defined by
\be
t = {\tilde{t} \over \alpha} \, , \qquad r = r_0+\alpha\tilde{x} \, ,
\ee
where $r_0=r_+=r_-=2\ell\sqrt{GM_0}$, the solutions
(\ref{metric0}-\ref{dilaton0}) have a well-defined limit when 
$\alpha\rightarrow 0$:
\bea
ds^2 &=& -\left( {4\tilde{x}^2 \over \ell^2}-a^2+{\cal O}(\alpha)\right)d\tilde{t}^2
+ \left( {4\tilde{x}^2 \over \ell^2}-a^2+{\cal O}(\alpha)\right) ^{-1}d\tilde{x}^2 
\, , \label{metric2} \\
\phi &=& {r_0 \over \ell}+{\alpha \over \ell}\tilde{x} \, , \label{dilaton2}
\eea
with $a^2=8M_0Gk$. This way we recover, in the $\alpha\rightarrow 0$ limit,
an AdS$_2$ geometry with curvature
$R=-{8 \over \ell^2}$. Arguing now as in the previous section and assuming
analogous boundary conditions for the asymptotic expansion of the 
two-dimensional AdS$_2$ metric
\bea
g_{\tilde{t}\tilde{t}} &=& -{4\tilde{x}^2 \over \ell^2}+\gamma_{\tilde{t}\tilde{t}}
+{\cal O}({1 \over \tilde{x}^2}) \, , \\
g_{\tilde{t}\tilde{x}} &=& {\gamma_{\tilde{t}\tilde{x}} \over \tilde{x}^3}
+{\cal O}({1 \over \tilde{x}^5}) \, , \\
g_{\tilde{x}\tilde{x}} &=& {\ell^2 \over 4\tilde{x}^2}
+{\gamma_{\tilde{x}\tilde{x}} \over \tilde{x}^4}
+{\cal O}({1 \over \tilde{x}^6}) \, ,
\eea
we find that 
\be
\Theta_{\tilde{t}\tilde{t}}=\kappa
(\gamma_{\tilde{t}\tilde{t}}-{\gamma_{\tilde{x}\tilde{x}} \over 2(\ell /2)^4})
\ee
transforms as
\be
\delta_{\epsilon}\Theta_{\tilde{t}\tilde{t}}=\epsilon (t)\Theta_{\tilde{t}\tilde{t}}'
+2\Theta_{\tilde{t}\tilde{t}}\epsilon'(t)
-{\kappa\ell^2 \over 4}\epsilon'''(t) \, . \label{law2}
\ee

Observe that the modification of the above expression with respect to the case
$J=0$ is due to the shift in the two-dimensional curvature, 
$R=-{8 \over \ell^2}$ instead of $R=-{2 \over \ell^2}$ for $J=0$.

  Since the AdS$_2$ codifies in some sense the black hole geometry in the 
near-extremal situation
\be
{M-M_0 \over M_0} = k\alpha^2\ll 1 \, ,
\ee
the idea now is to exploit this feature to explain the near-extremal entropy in 
terms of the asymptotic symmetries of AdS$_2$. To evaluate the central charge when
we approach to the extremal black hole we have to work out the Noether charges. 
The calculation is similar to that given in section 2, since only the derivative
term in the action (\ref{action3}) is relevant, and it coincides with that of
Jackiw-Teitelboim theory. Therefore, the Noether charges, to leading order in
$\alpha$, are
\be
K(\epsilon) = \epsilon(\tilde{t}){\alpha \ell \over 4G}{1 \over \ell}
\left(\gamma_{\tilde{t}\tilde{t}}
-{\gamma_{\tilde{x}\tilde{x}} \over 2(\ell /2)^4}\right) \, ,
\ee
and this implies that the coefficient $\kappa$ is
\be
\kappa = {\alpha \over 4G} \, .
\ee
Assuming $\tilde{t}$ varies in the interval $0<\tilde{t}<2\pi (R/2)^{-1/2} = \pi\ell$,
the central charge is
\be
c = 6 \kappa\ell = {3 \ell \over 2G}\alpha \, . \label{centralc2}
\ee

  We also want to compute the value of $L_0^R$ in the near-extremal black hole
solutions. Since $L_0^R = {\ell \over 2} K \quad (\epsilon(\tilde{t}) = 1)$ we find that
\be
L_0^R = {\ell \over 2}{\alpha \over 4G}{1 \over 2}a^2 = {1 \over 2}M_0k\alpha\ell \, . 
\label{eigenvalue} 
\ee
  It is interesting to observe that, with respect to the time 
$t={\tilde{t} \over \alpha}$, the generators $L_{m(\tilde{t})}^R$ shift into
$L_{m(t)}^R=\alpha L_{m(\tilde{t})}^R$, and then $L_{0(t)}^R$ is equal to 
one half of the 
deviation of the mass from the extremal case:
\be
L_{0(t)}^R={1 \over 2}M_0\ell k\alpha^2 = {1 \over 2}(M-M_0)\ell \, .
\ee

  From (\ref{centralc2}) and (\ref{eigenvalue})
we can evaluate, via Cardy's formula, the degeneracy of
states if $M-M_0$ is large in the microscopic sense
\be
2\pi\sqrt{cL_0^R \over 6} = \pi\sqrt{\ell^2(M-M_0) \over 2G} \, , \label{entropy4}
\ee
which turns out to be just the difference between the entropy of a nearly extremal black
hole and a extremal one
\be
\Delta S = S-S_e=\pi\sqrt{\ell^2(M-M_0) \over 2G} \, .
\ee
Therefore, the statistical entropy (\ref{entropy4})
 just account for microscopic excitations from the
extremal macroscopic state.

\section{Near-extremal Reissner-Nordstr\"om black holes}

  Let  us start with the Einstein-Hilbert action in 3+1 dimensions
\be 
{1 \over 16\pi G}\int d^4x\sqrt{-g^{(4)}}(R^{(4)}-G(F^{(4)})^2) \, .
\label{HEaction}
\ee
Imposing spherical symmetry on the electromagnetic field and on the metric
\be
ds_{(4)}^2 = g_{\mu\nu}^{RN}dx^{\mu}dx^{\nu}
+{1 \over 2}\ell^2\bar{\phi}^2(x)d\Omega^2 \, ,
\ee
where $d\Omega^2$ is the metric on the two-sphere and $\ell$ is the Planck
length ($\ell^2=G$), the action (\ref{HEaction}) reduces to
\be
\int d^2x\sqrt{-g}\left[{1 \over 2}\left({\bar{\phi}^2 \over 4}R
+{1 \over 2}\mid\nabla\bar{\phi}\mid^2+{1 \over \ell^2}\right)
-{\ell^2 \over 8}\bar{\phi}^2F^{\mu\nu}F_{\mu\nu}\right] \, .
\label{actionred}
\ee
To perform a similar analysis to that of section 3 we need to reparametrise
the fields to eliminate the kinetic term in (\ref{actionred}) and bring
the action to the more reduced form (\ref{action2}). To this end we introduce
the new fields
\bea
\phi &=& {\bar{\phi}^2 \over 4} \, , \label{resc1} \\
g_{\mu\nu} &=& \sqrt{2\phi}g_{\mu\nu}^{RN} \, . \label{resc2}
\eea
The two-dimensional effective action becomes
\be
{1 \over 2}\int d^2x\sqrt{-g}\phi\left( R+{1 \over \sqrt{2}\ell^2\phi^{3/2}}
-\sqrt{2}\phi^{1/2}\ell^2F^{\mu\nu}F_{\mu\nu}\right) \label{action4}
\ee
and the equations of motion of the electromagnetic field yield to
\be
{4\sqrt{2}\ell\phi^{3/2} \over \sqrt{-g}}F_{+-} = Q \, , \label{const2}
\ee
where $Q$ is an integration constant. Plugging (\ref{const2}) into
(\ref{action4}) we get
\be
{1 \over 2}\int d^2x\sqrt{-g}(\phi R+V(\phi)) \, ,
\ee
where
\be
V(\phi) = {1 \over \ell^2}\left({1 \over \sqrt{2\phi}}
-{\ell^2 Q^2 \over (2\phi)^{3/2}}\right) \, .
\ee
  The general solution with a non-constant dilaton is
\bea
ds^2 &=& -(\sqrt{2\phi}+{\ell^2 Q^2 \over \sqrt{2\phi}}-2M\ell)dt^2
+(\sqrt{2\phi}+{\ell^2 Q^2 \over \sqrt{2\phi}}-2M\ell)^{-1}dx^2 \quad
\label{metric5} \\
\phi &=& {x \over \ell} \, . \label{dilaton5}
\eea
Note that the rescaling (\ref{resc1}-\ref{resc2}) map the above solutions into
the standard form
\bea
(ds^2)^{RN} &=& -\left( 1-{2GM \over r}+{Q^2G^2 \over r^2}\right) dt^2 \\
&+&\left( 1-{2GM \over r}+{Q^2G^2 \over r^2}\right)^{-1}dr^2 \\
\bar{\phi} &=& \sqrt{2}{r \over \ell} \, .
\eea
As is well known, there are two event horizons, located at
\be
\sqrt{2\phi} = \ell (M\pm\sqrt{M^2-Q^2}) \, .
\ee
Perturbing the solution (\ref{metric5}) around the degenerate radius 
$x_0={1 \over 2}\ell^3M_0^2$ of the extremal solution $M_0=\mid Q\mid$
\bea
M &=& M_0(1+k\alpha^2) \, , \\
t &=& {\tilde{t} \over \alpha}\, , \qquad x = x_0+\alpha\tilde{x} \, , 
\eea
we get in the near-horizon limit $\alpha\rightarrow 0$
\be
ds^2 = -\left({1 \over \ell^5\mid Q\mid^3}\tilde{x}^2-2\mid Q\mid k\ell\right)
d\tilde{t}^2
+\left({1 \over \ell^5\mid Q\mid^3}\tilde{x}^2-2\mid Q\mid k\ell\right)^{-1}d\tilde{x}^2
\, .
\ee
So, the curvature is $R_0={2 \over \ell^5M_0^3}$ and $a^2=4M_0\ell$.
Note that for the metric $g_{\mu\nu}^{RN}$ the curvature is ${2 \over \ell^4 M_0^2}$,
which corresponds to that of the Robinson-Bertotti geometry.

  Proceeding in a parallel way as in the case of near-extremal BTZ black holes,
we find here that the Noether charges are
\be
K(\epsilon) = \epsilon(\tilde{t}){\alpha \over \ell}\left(\gamma_{\tilde{t}\tilde{t}}-
{1 \over 2\ell^{10}Q^6}\gamma_{\tilde{x}\tilde{x}}\right) \, ,
\ee
and these yield to the central charge
\be
c = 12\mid Q\mid^3{\ell^4\alpha \over \beta}
\ee
if $\tilde{t}\in [0, 2\pi\beta]$.
The value of $L_0^R$ near extremality is
\be
L_0^R = \mid Q\mid k\alpha \beta
\ee
and using Cardy's formula we obtain
\be
\Delta S = 2\pi\sqrt{cL_0^R \over 6} = 2\pi\sqrt{2Q^3\ell^4\Delta M} \, , \label{varS}
\ee
where
\be
\Delta M = \mid Q\mid k\alpha^2 = M-M_0 \, .
\ee
It is now easy to see that the statistical expression (\ref{varS})
exactly agrees with the
deviation of the Bekenstein-Hawking
entropy of near-extremal black holes from the extremal case
$S_e = \pi Q^2\ell^2$
\be
S = \pi\ell^2(\mid Q\mid + \Delta M + \sqrt{2\mid Q\mid\Delta M+(\Delta M)^2})^2
= S_e + \Delta S + {\cal O}((\Delta M)^{3/2}) \, .
\ee

\section{AdS$_2$/CFT$_1$ correspondence and near-extremal black holes}

  In this section we shall generalise the argument leading to the statistical
explanation of the near-extremal Bekenstein-Hawking entropy of BTZ and 
Reissner-Nordstr\"om black holes to a wider family of black holes. We shall
consider a generic black hole solution, in an arbitrary dimension n, which
can be described by the metric
\be
ds^2 = g_{\mu\nu}dx^{\mu}dx^{\nu}+{1 \over 2}\ell^2\phi^2d\Omega^{n-2}
\ee
where $\ell$ is the Planck length of the theory. By dimensional reduction and 
integrating the equations of motion of any abelian gauge field we 
can arrive at an effective
two-dimensional theory. An additional conformal rescaling of the metric and a redefinition
of the dilaton field yield into an action of the form \cite{BL, GKL}
\be
{1 \over 2G}\int d^2x\sqrt{-g}(R\phi+\ell^{-2}V(\phi))  \, , \label{action5}
\ee
where $V(\phi)$ is a potential function parametrising the original theory and G is
a dimensionless constant playing the role of Newton constant in two-dimensions.
The solutions for the 2D effective metric are
\bea
ds^2 &=& -(J(\phi)-2M\ell)dt^2+(J(\phi)-2M\ell)^{-1}dr^2 \, , \label{metric6} \\
\phi &=& {r \over \ell} \, ,
\eea
where $V'(\phi)=J(\phi)$. The horizons are the solutions of the equation 
$J(\phi)=2M\ell$ and we have a degeneration at the zeros of the potential
\be
V(\phi_0)=J'(\phi_0)=0 \, .
\ee
  If we perturb around the degenerate radius of coincident horizons
\bea
M &=& M_0(1+k\alpha^2) \, , \\
t &=& {\tilde{t} \over \alpha} \, , \\
r &=& r_0+\alpha\tilde{x}\, ,
\eea
the two-dimensional metric transforms into
\be
ds^2 = -(-{R_0 \over 2}\tilde{x}^2-2kM_0\ell+{\cal O}(\alpha))d\tilde{t}^2
+(-{R_0 \over 2}\tilde{x}^2-2kM_0\ell+{\cal O}(\alpha))^{-1}d\tilde{x}^2 \, ,
\ee
where
\be
R_0 = {J''(\phi_0) \over \ell^2} \, . \label{R_0}
\ee
  Imposing boundary conditions of the form
\bea
g_{\tilde{t}\tilde{t}} &=& {R_0 \over 2}\tilde{x}^2+\gamma_{\tilde{t}\tilde{t}}
+... \, , \\
g_{\tilde{t}\tilde{x}} &=& {\gamma_{\tilde{t}\tilde{x}} \over \tilde{x}^3} \, , \\
g_{\tilde{x}\tilde{x}} &=& -{2 \over R_0}{1 \over \tilde{x}^2}
+{\gamma_{\tilde{x}\tilde{x}} \over \tilde{x}^4}+... \, , \\
\eea
and working in the gauge $\gamma_{\tilde{t}\tilde{x}}=0$, the Noether charges can be
worked out without difficulty because of the simple form of the two-dimensional
effective action (\ref{action5}).
\be
K(\epsilon) = \epsilon(\tilde{t})\Theta_{\tilde{t}\tilde{t}} \, , 
\ee
where $\Theta_{\tilde{t}\tilde{t}}$ is the stress tensor
\be
\Theta_{\tilde{t}\tilde{t}} = {\alpha \over \ell G}\left( \gamma_{\tilde{t}\tilde{t}}
-{1 \over 2}({R_0 \over 2})^2\gamma_{\tilde{x}\tilde{x}}\right) \, .
\ee
Assuming a periodicity of $2\pi\beta$ in $\tilde{t}$, we obtain
\bea
c &=& {24 \alpha \over \ell G R_0\beta} \, , \\
L_0^R &=& {M_0 k\alpha\beta \over G} \, .
\eea
Applying now Cardy's formula we get
\be
\Delta S = 2\pi\sqrt{4M_0\ell k\alpha^2 \over R_0\ell^2G^2} \label{varS2}
\ee
and, taking into account (\ref{R_0})and that
\be
M_0\ell k\alpha^2 = (M-M_0)\ell = {1 \over 2}(J(\phi_h)-J(\phi_0)) \, ,
\ee
where $\phi_h$ is the value of the dilaton at the outer horizon, we can rewrite
(\ref{varS2}) as
\be
\Delta S = {2\pi \over G}\sqrt{2M_0(J(\phi_h)-J(\phi_0)) \over J''(\phi_0)} \, .
\label{entropy6} 
\ee
On the other hand,
the Bekenstein-Hawking entropy for the two-dimensional effective theory is given by
the simple expression \cite{GKL}
\be
S={2\pi \over G}\phi_h \, ,
\ee
and therefore,
\be
\Delta S = {2\pi \over G}(\phi_h-\phi_0) \, . \label{entropy7}
\ee
  Expanding $J(\phi)$ around the extremal situation ($J'(\phi_0)=0$)
\be
J(\phi_h)=J(\phi_0)+{1 \over 2}J''(\phi_0)(\phi_h-\phi_0)^2+...
\ee
and, in the near-extremal approximation, we have
\be
\frac{2(J(\phi_h)-J(\phi_0))}{J''(\phi_0)} = (\phi_h-\phi_0)^2 \, ,
\ee
implying the equality between the statistical expression (\ref{entropy6})
 and the thermodynamical one (\ref{entropy7}).

\section{Conclusions and final remarks}
  We have shown that the asymptotic symmetries of BTZ and Reissner-Nordstr\"om extremal
black holes, whose near-horizon geometry is AdS$_2\times$S$^n$ (n=1,2 respectively)
are powerful enough to control the deviation of the Bekenstein-Hawking entropy of 
nearly extremal black holes from the extremal situation. We have also argued that the
above results can be generalised for arbitrary black holes near extremality if they can be
described by an effective two-dimensional dilaton gravity theory. 

Our approach is based on a realization of the boundary conformal field theory in terms of
the sub-leading terms in the asymptotic expansion of the metric field. The evaluation
of the Noether charges associated with the asymptotic symmetries near extremality allows
to compute the central charge and the value of $L_0^R$. These values depend on an
arbitrary parameter $\beta$ in such a way that $cL_0^R$, and hence the statistical 
entropy, has an absolute meaning \footnote{A similar situation appears in
\cite{Carlip3, Solod}}. However, in the present context the physical excitations are associated
to the "would-be gauge" diffeomorphisms characterised by the functions
$\epsilon(\tilde{t})$ and these degrees of freedom have an effective central charge
$c_{eff}=1$ (see \cite{Nav2}). Therefore, it could appear natural to choose $\beta$
in such a way that $c=1$ and bypass the question of the discrepancy between $c$
and $c_{eff}$. This type of argument was put forward in 2+1 gravity in \cite{Bb,BBBB}
and could have some unexpected consequences.

\section*{Acknowledgements}
This research has been partially supported by the 
Comisi\'on Interministerial de Ciencia y Tecnolog\'{\i}a 
and DGICYT.
P. Navarro acknowledges the Ministerio de Educaci\'on y Cultura for a FPU 
fellowship.
We want to thank J.Cruz and D.J.Navarro for useful conversations.

\end{document}